\documentclass[epj-spec]{svjour}
\usepackage[T1]{fontenc}
\usepackage{lmodern}
\usepackage{float}
\usepackage{amsmath}
\usepackage{graphicx}
\usepackage{esint}

\makeatletter

\usepackage{graphics}

\newcommand{\dk}{\mathrm{d}\hspace{-.5pt}k}
\newcommand{\dw}{\mathrm{d}\omega}

\newcommand{\ddk}{\mathrm{d}^{3}k}
\newcommand{\dvk}{\mathrm{d}^{4}k}
\newcommand{\dnk}{\mathrm{d}^{n}k}
\newcommand{\nb}{n_\text{B}}
\newcommand{\nf}{n_\text{F}}
\newcommand{\PiT}{\Pi_\text{T}}
\newcommand{\PiL}{\Pi_\text{L}}
\newcommand{\DT}{D_\text{T}}
\newcommand{\DL}{D_\text{L}}

\newcommand{\I}{\text{Im}}
\newcommand{\R}{\text{Re}}
\newcommand{\e}{\varepsilon}

\newcommand{\MeV}{\text{M}\hspace{-.5px}e\hspace{-.8px}\text{V}}

\makeatother

\begin{document}

\title{Plasmons, plasminos and Landau damping in a quasiparticle model of
the quark-gluon plasma}

\author{R.~Schulze\thanks{\email{r.schulze@fzd.de}}
\and M.~Bluhm
\and B.~K\"ampfer}

\institute{Forschungszentrum Dresden-Rossendorf, PF 510119, 01314 Dresden, Germany\and Institut
f\"ur Theoretische Physik, TU Dresden, 01062 Dresden, Germany}

\abstract{A phenomenological quasiparticle model is surveyed for 2+1 quark
flavors and compared with recent lattice QCD results. Emphasis is
devoted to the effects of plasmons, plasminos and Landau damping.
It is shown that thermodynamic bulk quantities, known at zero chemical
potential, can uniquely be mapped towards nonzero chemical potential
by means of a thermodynamic consistency condition and a stationarity
condition.}

\maketitle
\titlerunning{Towards an EOS for the cold and dense QGP}

\authorrunning{R.~Schulze et al.}

\section{Introduction}

\label{intro}

Intense experimental and theoretical investigations \cite{KMR03}
suggest the existence of a new, deconfined phase of strongly interacting
matter, where quarks and gluons form a fluid or gas, the quark-gluon
plasma (QGP). If confirmed, the QGP would have existed during the
Big Bang prior to hadronization and might be found inside of massive
neutron stars. Indeed, recent results from the Relativistic Heavy
Ion Collider (RHIC) experiments point to the formation of a quark-gluon
medium of low viscosity \cite{BRA05,PHO05,STA05,PHE05}.

However, on the theoretical side, much work remains to be done. Perturbative
solutions of QCD \cite{AZ95,ZK95,Kaj03,Vuo03a,Vuo03b,IRV04} are limited
to the region of asymptotic freedom and fail for the strong coupling
regime (e.g.~in the vicinity of the pseudocritical temperature \cite{BI02} of
deconfinement $T_{c}$; at somewhat higher temperatures, say above $2T_c$
resummation improves the convergence of perturbation theory noticeably
\cite{KPP97,ABS99,BIR01}). Numerical evaluations of the full theory,
on the other hand, are still restricted to small chemical potential
as being useful for the Big Bang or heavy-ion collisions at present
RHIC top energies or future LHC energies. However, at RHIC bottom energies,
at SPS energies and, in particular, at FAIR energies, baryon density
effects become significant and require different approaches.

Quasiparticle models (QPM), describing the quark-gluon plasma as assembly
of essentially non-interacting excitations emerging from the strong
interaction, have proven to represent useful phenomenological parametrizations
of QCD thermodynamics above $T_{c}$ \cite{Pes94,LH98,Pes02,LR03,TSW04}.
At zero chemical potential, lattice results are described with surprising
accuracy allowing the adjustment of model parameters. Thermodynamic
self-consistency, supplemented by the stationarity of the thermodynamic
potential, can then be used to extrapolate thermodynamic properties
of systems to nonzero net baryon density.

Our quasiparticle model \cite{KBS06,BKS06,BKS07a,BKS07b} is based
on the HTL approximation \cite{BP90a} to the 1-loop self-energies.
This gives rise to four quasiparticle families. While quasiquarks
and transversal gluons within the model represent excitations with
quantum numbers of actual quarks and gluons with modified masses,
quark holes (plasminos) and longitudinal gluons (plasmons) are quanta
of collective excitations. The residues of the poles in the spectral
density of the collective modes vanish exponentially for momenta $k\sim T$,
from which the main contributions to thermodynamic phase space integrals
originate. Therefore, they were neglected in the previous simple form
of the model (dubbed eQP in \cite{Pes05}). Additionally, damping
contributions were neglected as they are small at zero chemical potential,
$\mu=0$.

The procedure of mapping the eQP results from $\mu=0$ into the $T$-$\mu$
plane is plagued by some ambiguities leading to non-unique solutions
close to the presumed phase transition. Therefore, the model is restricted
to sufficiently large temperatures. First attempts to include collective
excitations into a two-flavor quasiparticle model at nonzero chemical
potential have been made \cite{RR03} and suggest that these ambiguities
might vanish. In this work we show that both collective excitations
and damping effects are necessary to preserve the self-consistency
of the model and ensure unique solutions when extrapolating towards
large baryon densities at moderate temperatures. In the present work,
the 2+1 flavor case is considered, allowing the use and extrapolation
of recent lattice data \cite{Kar07}. 

Our paper is organized as follows. Section \ref{sec:derivation} will
comprise the derivation of the full HTL-based QPM, including collective
modes and damping, as a series of approximations from QCD. The necessary
further approximations leading to the eQP and its problems are discussed
in section \ref{sec:eQP}. The results for both models are then contrasted
in section \ref{sec:fullHTL} and investigated in some detail. Finally,
a conclusion is given in section \ref{sec:conclusion}.

\section{Derivation of the full HTL model\label{sec:derivation}}

\subsection{The effective action}

A connection of the fundamental theory of QCD and the thermodynamic
potential of the QGP is provided by the Luttinger-Ward formalism \cite{LW60,Bay62}
as shown in \cite{BKS07a}. Alternatively, the Cornwall-Jackiw-Tomboulis
(CJT) formalism \cite{CJT74} may be used, as a translationally invariant
QGP in equilibrium and without spontaneously broken symmetries is
considered. In this case, both formalisms are equivalent \cite{Ris03}.

The CJT formalism requires the stationarity of the effective action\begin{align}
\Gamma[D,S]=I & -\frac{1}{2}\left\{ \text{Tr}\left[\ln D^{-1}\right]+\text{Tr}\left[D_{0}^{-1}D-1\right]\right\} \nonumber \\
 & +\quad\left\{ \text{Tr}\left[\ln S^{-1}\right]+\text{Tr}\left[S_{0}^{-1}S-1\right]\right\} \,\,+\,\,\Gamma_{2}[D,S],\label{eq:cjt effective action}\end{align}
where $I$ is the classical action and $D$ and $S$ are the full
gluon and quark propagators while the subscript $0$ denotes the respective
free equivalents. The functional $\Gamma_{2}$ represents the sum
over all two-particle irreducible skeleton graphs of the theory. The
traces $\text{Tr}$ contain the integration over the four-dimensional
phase space as well as a trace $\text{tr}$ over discrete indices.
The integration is performed using the imaginary time formalism \cite{LeB96,Kap89,YHM95}.
For the grand canonical potential $\Omega=-T\Gamma$ \cite{Bro92,Riv88}
this yields\begin{eqnarray}
\frac{\Omega}{V} & = & \mbox{tr}\!\!\int\!\!\frac{\dvk}{(2\pi)^{4}}\nb(\omega)\,\mbox{Im}\!\left(\ln D^{-1}-\Pi D\right)+2\,\mbox{tr}\!\!\int\!\!\frac{\dvk}{(2\pi)^{4}}\nf(\omega)\,\mbox{Im}\!\left(\ln S^{-1}-\Sigma S\right)-\frac{T}{V}\Gamma_{2},\label{eq:Omegafinal}\end{eqnarray}
where $\nb=(\exp\,(\beta\omega)-1)^{-1}$ with $\beta=1/T$ is the
Bose-Einstein, and $\nf=(\exp(\beta(\omega-\mu))+1)^{-1}$ the Fermi-Dirac
distribution function.

\subsection{Application to QCD}

In our present approach, the infinite sum $\Gamma_{2}$ is truncated
at 2-loop order leaving the contributions exhibited e.g.~in equation
(25) in \cite{BKS07a}. Within the CJT formalism, the self-energies
then follow from a functional derivative of $\Gamma_{2}$, giving
the well-known 1-loop self-energies (e.g.~equations (26) and (27)
ibid.). In order to achieve a gauge invariant formulation of the model,
we apply an additional approximation of hard thermal loops (HTL).
Although originally being derived for soft external momenta $\omega,k\sim gT\ll T$,
HTL results coincide with the complete one-loop results on the lightcone
\cite{BKS07a,Pes98b} and thus provide the correct limiting behaviour.

The resulting HTL self-energies can be found in textbooks. Here we
follow the conventions of Blaizot et al.~\cite{BIR01}, where essential
features of our model have been worked out, and use\begin{eqnarray}
\Pi_{\mu\nu} & = & \PiT(\omega,k)\left(\Lambda_{\text{T}}(\vec{k})\right)_{\mu\nu}-\,\,\,\,\PiL(\omega,k)\left(\Lambda_{\text{L}}(\vec{k})\right)_{\mu\nu},\\
\gamma_{0}\Sigma & = & \Sigma_{+}(\omega,k)\,\,\,\,\,\Lambda_{+}(\vec{k})\quad-\,\,\,\,\Sigma_{-}(\omega,k)\,\,\,\,\Lambda_{-}(\vec{k})\end{eqnarray}
with the scalar self-energies\begin{eqnarray}
\PiT(\omega,k)=\frac{m_{D}^{2}}{2}\left(1+\frac{\omega^{2}-k^{2}}{k^{2}}\PiL(\omega,k)\right), &  & \PiL(\omega,k)=m_{D}^{2}\left(1-\frac{\omega}{2k}\ln\frac{\omega+k}{\omega-k}\right),\label{eq:HTL PiT}\\
\text{and}\quad\quad\Sigma_{\pm}(\omega,k) & = & \frac{\hat{M}^{2}}{k}\left(1-\frac{\omega\mp k}{2k}\ln\frac{\omega+k}{\omega-k}\right),\label{eq:HTL Sigmapm}\end{eqnarray}
where $\hat{M}(T,\,\mu,\, g^{2})$ is the thermal fermion
mass or plasma frequency and $m_{D}(T,\,\mu,\, g^{2})$ denotes
the Debye screening mass\begin{eqnarray}
m_{D}^{2}=\!\left(\left[2N_{c}\!+\! N_{q}\!+\! N_{s}\right]T^{2}+\frac{N_{c}}{\pi^{2}}\sum_{i}\mu_{i}\right)\frac{g^{2}}{6} & \quad\text{and}\quad & \hat{M}^{2}=\frac{N_{c}^{2}-1}{16N_{c}}\left(T^{2}+\frac{\mu^{2}}{\pi^{2}}\right)g^{2}.\label{eq:mDebye and plasma freq}\end{eqnarray}
The number of colors $N_{c}$ is fixed at $3$. The numbers of light
quarks $N_{q}=2$ and one strange quark, $N_{s}=1$, are chosen as
in the lattice calculations \cite{Kar07}.

\subsection{Properties of HTL self-energies and dispersion relations}

\begin{figure}[t]
\noindent \begin{centering}
\includegraphics[scale=0.7]{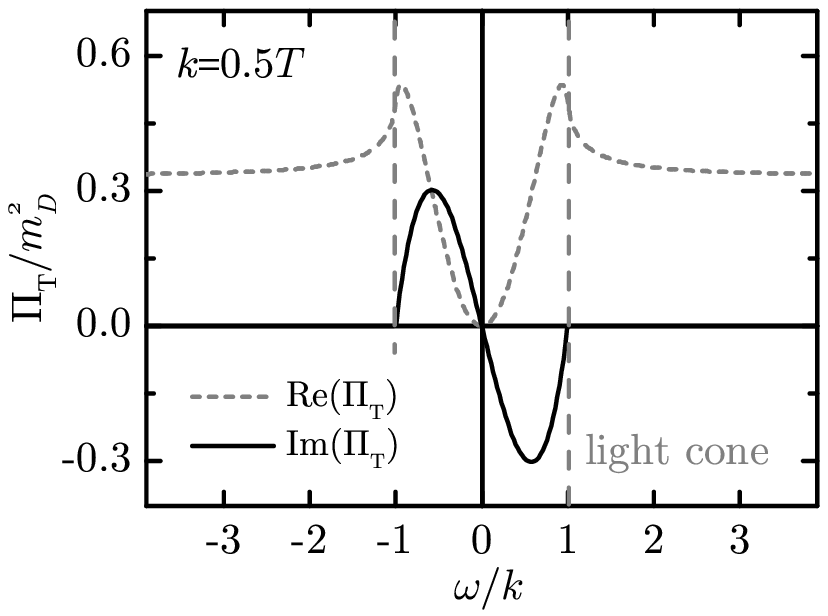}~~~~~~~~\includegraphics[scale=0.7]{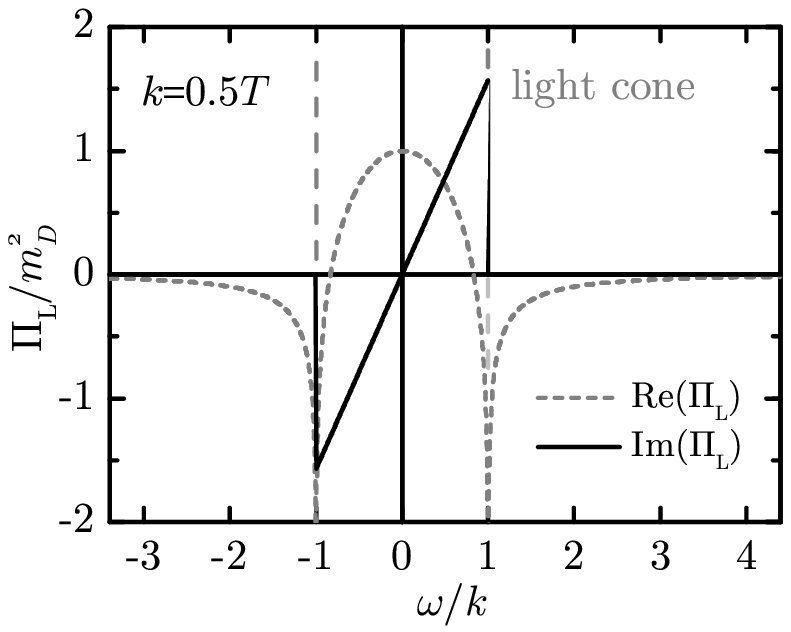}
\par\end{centering}

\caption{The real and imaginary parts of the retarded transverse (left) and
longitudinal (right) gluon self-energies scaled by the Debye mass
squared are shown as functions of the energy $\omega$ scaled by the
momentum $k$ which is fixed at $k=0.5T$. \label{fig:gluon se}}

\end{figure}

\begin{figure}
\noindent \begin{centering}
\includegraphics[scale=0.7]{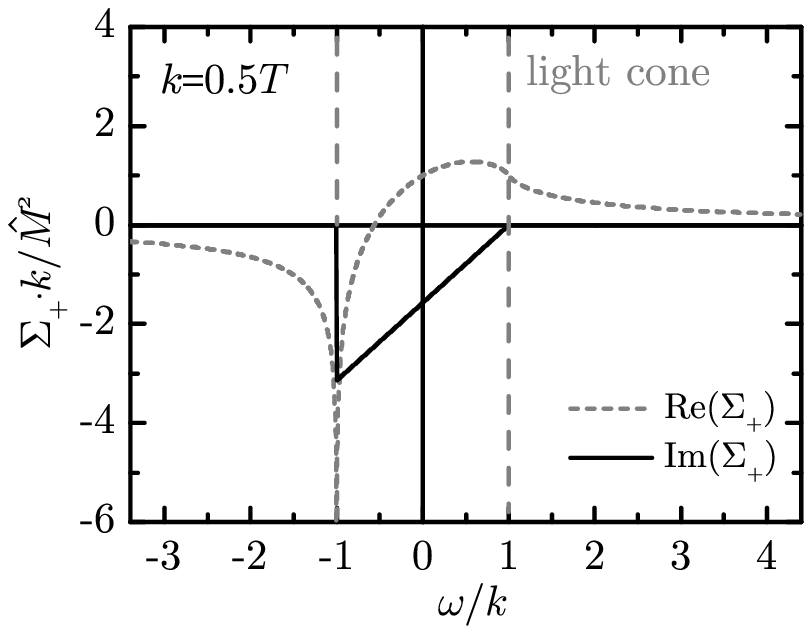}~~~~~~~~\includegraphics[scale=0.7]{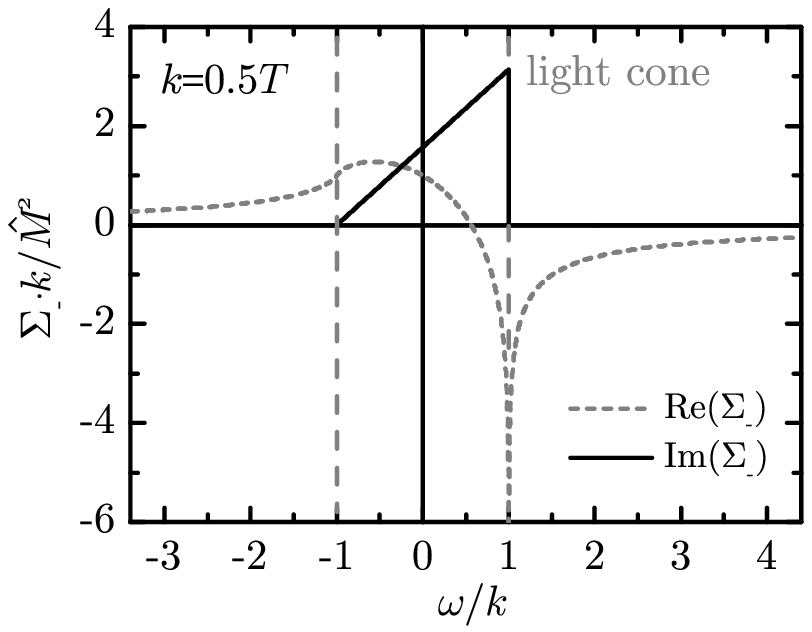}
\par\end{centering}

\caption{The real and imaginary parts of the retarded quark self-energies for
the normal (left) and abnormal branch (right) scaled by the plasma
frequency squared are shown as functions of the energy $\omega$ scaled
by the momentum $k$ which is fixed at $k=0.5T$. \label{fig:quark se}}

\end{figure}

The real and the imaginary parts of the HTL self-energies (\ref{eq:HTL PiT})
and (\ref{eq:HTL Sigmapm}) are \cite{LeB96}\begin{eqnarray}
\R\Pi_{T}=\frac{m_{D}^{2}}{2}\left(1+\frac{\omega^{2}\!-\! k^{2}}{k^{2}}\R\PiL(\omega,k)\right)\!, &  & \I\Pi_{T}=\frac{1}{2}m_{D}^{2}\frac{\omega^{2}\!-\! k^{2}}{k^{2}}\frac{\omega}{2k}\pi\Theta\left(k^{2}\!-\!\omega^{2}\right)\varepsilon(k),\label{eq:Im PiT}\\
\R\Pi_{L}=m_{D}^{2}\left(1-\frac{\omega}{2k}\ln\left|\frac{\omega+k}{\omega-k}\right|\right)\!, &  & \I\Pi_{L}=m_{D}^{2}\frac{\omega}{2k}\pi\Theta\left(k^{2}-\omega^{2}\right)\varepsilon(k),\label{eq:Im PiL}\\
\R\Sigma_{\pm}=\frac{\hat{M}^{2}}{k}\left(1-\frac{\omega\mp k}{2k}\ln\left|\frac{\omega+k}{\omega-k}\right|\right)\!, &  & \I\Sigma_{\pm}=\frac{\hat{M}^{2}}{k}\frac{\omega\mp k}{2k}\pi\Theta\left(k^{2}-\omega^{2}\right)\varepsilon(k),\label{eq:Im Sigmapm}\end{eqnarray}
where $\e(k)$ is the sign function. The real parts are symmetric
with respect to $\omega$, while the imaginary parts are antisymmetric
and differ from zero only for $|\omega|<k$, i.e.~below the light
cone. The real and imaginary parts for $k=0.5T$ are shown in Figure
\ref{fig:gluon se}. Analogously, the quark self-energies fulfill
the parity relations $\R\Sigma_{+}(-\omega)=\R\Sigma_{-}(\omega)$
and $\I\Sigma_{+}(-\omega)=-\I\Sigma_{-}(\omega)$ as shown for $k=0.5T$
in Figure \ref{fig:quark se}.%

It follows directly from eqs.~(\ref{eq:Im PiT})-(\ref{eq:Im Sigmapm})
that the HTL self-energies do not account for quasiparticle widths
since the imaginary parts are zero at the poles of the quasiparticle
propagators, i.e.~above the light cone. The nonzero imaginary parts
of the self-energies below the light cone are due to \emph{Landau
damping} (LD). LD is a collective effect caused by energy transfer
between the gauge field and plasma particles with velocities close
to the phase velocity ({}``resonant particles'').

Even though the imaginary parts are formally nonzero only below the
light cone, retardation leads to an infinitely small contribution
even above the light cone, giving a definite sign to the self-energies
for all energies: $\e(\I\PiT(\omega))=-\e(\omega)$, $\e(\I\PiL(\omega))=+\e(\omega)$
and $\e(\I\Sigma_{\pm}(\omega))=\mp1$. Note that this is not related
to Landau damping which is found below the light cone only.

The HTL propagators follow from Dyson's equations as $\DT^{-1}=-\omega^{2}+k^{2}+\PiT$,
$\DL^{-1}=-k^{2}-\PiL$ and $S_{\pm}^{-1}=-\omega\pm(k+\Sigma_{\pm})$.
On-shell (quasi)particles satisfy a dispersion relation determined
by $\R D_{\text{T},\text{L}}^{-1}=0$ and $\R S_{\pm}^{-1}=0$ respectively.
It is, therefore, useful to first investigate the real part of the
inverse retarded HTL propagators.%
\begin{figure}[t]
\noindent \begin{centering}
\includegraphics[scale=0.7]{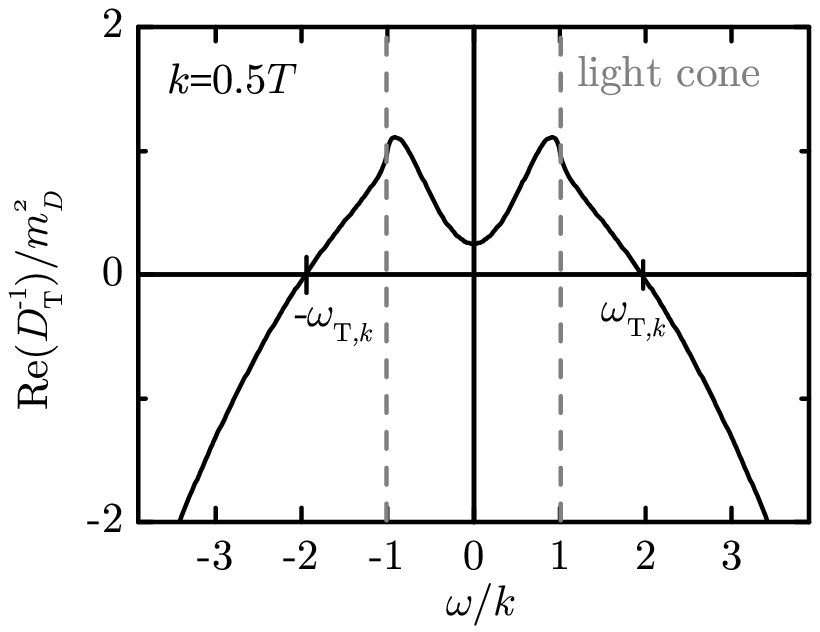}~~~~~~~~\includegraphics[scale=0.7]{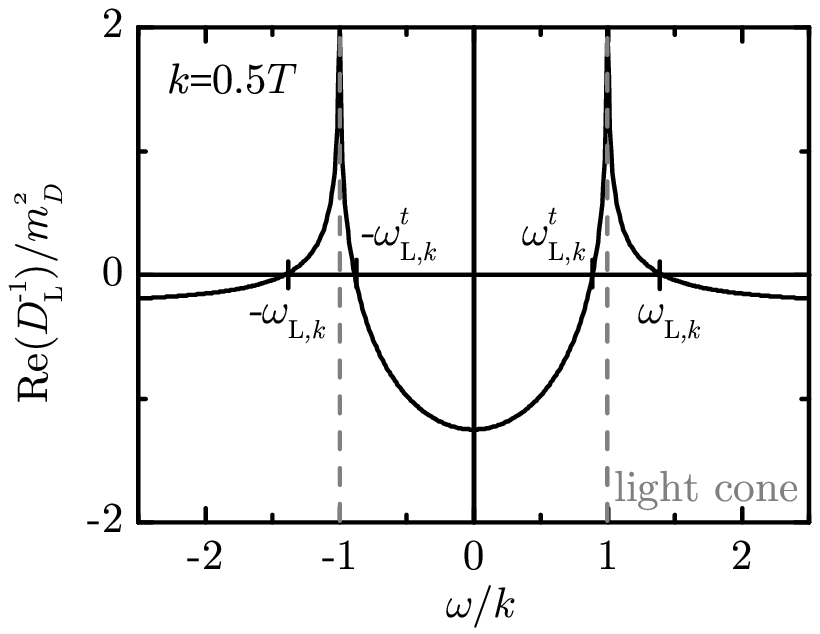}
\par\end{centering}

\caption{The real parts of the inverse gluon propagators $D_{\text{T,L}}^{-1}$
scaled by the Debye screening mass squared are shown as functions
of the energy $\omega$ scaled by the momentum $k$ which is fixed
at $k=0.5T$. \label{fig:ReGluPropm1}}

\end{figure}
\begin{figure}[t]
\noindent \begin{centering}
\includegraphics[scale=0.7]{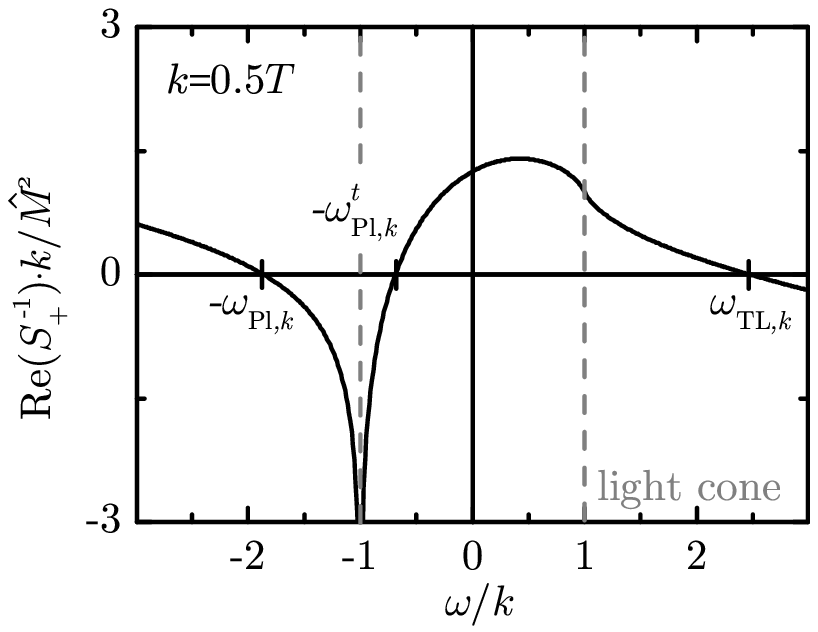}~~~~~~~~\includegraphics[scale=0.7]{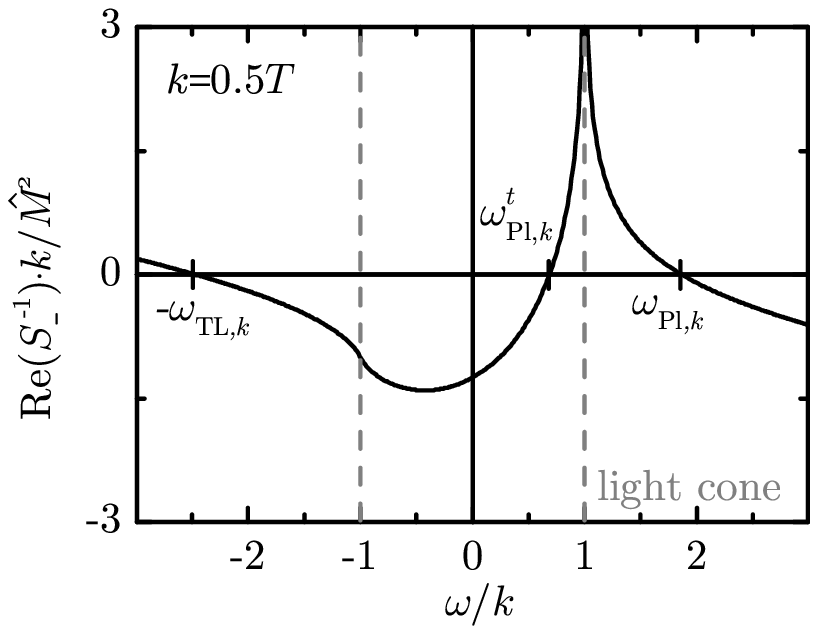}
\par\end{centering}

\caption{The real parts of the inverse quark propagators $S_{\pm}^{-1}$ scaled
by the fermionic mass parameter squared are shown as functions of
the energy $\omega$ scaled by the momentum $k$ which is fixed at
$k=0.5T$. \label{fig:ReQuarkPropm1}}

\end{figure}

Due to symmetry properties of $D_{\text{T,L}}^{-1}$ there is just
one positive-energy dispersion relation above the light cone: $\omega_{\text{T},k}$
and $\omega_{\text{L},k}$, respectively. This means that - up to
the sign - transverse and longitudinal gluons have the same dispersion
relations as their anti(quasi)particle counterparts. The additional
tachyonic dispersion relation for longitudinal gluons is related to
Landau damping. Figure \ref{fig:ReGluPropm1} explicitly shows the
real parts for fixed momentum $k=0.5T$. Both $D_{\text{T},\text{L}}^{-1}$
are symmetric with respect to $\omega$. The zero of $\R\DT^{-1}$
determines the dispersion relation $\omega_{\text{T},k}$ for transverse
gluons. The zero of $\R\DL^{-1}$ above the light cone indicates the
dispersion relation $\omega_{\text{L},k}$ of longitudinal gluons,
while the tachyonic dispersion relation $\omega_{\text{L},k}^{t}$
(below the light cone) is due to Landau damping.

The inverse quark propagators are not symmetric but, as a consequence
of the symmetry of the self-energy, satisfy the parity property $\R S_{+}^{-1}(-\omega)=-\R S_{-}^{-1}(\omega)$
(cf.~Figure \ref{fig:ReQuarkPropm1}). Hence, quarks are described
by the positive energy dispersion relation related to $S_{+}$, while
the dispersion relation of antiquarks is found from the negative energy
solution of $\R S_{-}^{-1}=0$. The remaining two dispersion relations
represent collective quark excitations: the positive energy dispersion
relation related to $S_{-}$ describes the plasminos, while the negative
energy solution of $\R S_{+}^{-1}=0$ represents antiplasminos. Again,
a tachyonic solution appears within the regime of Landau damping.

\begin{figure}[t]
\noindent \begin{centering}
\includegraphics[scale=0.7]{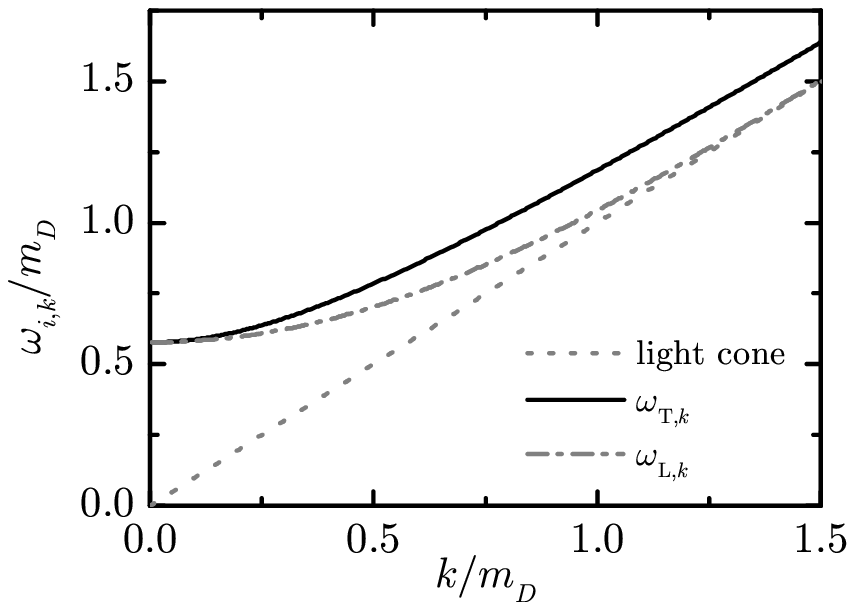}~~~~~~~~\includegraphics[scale=0.7]{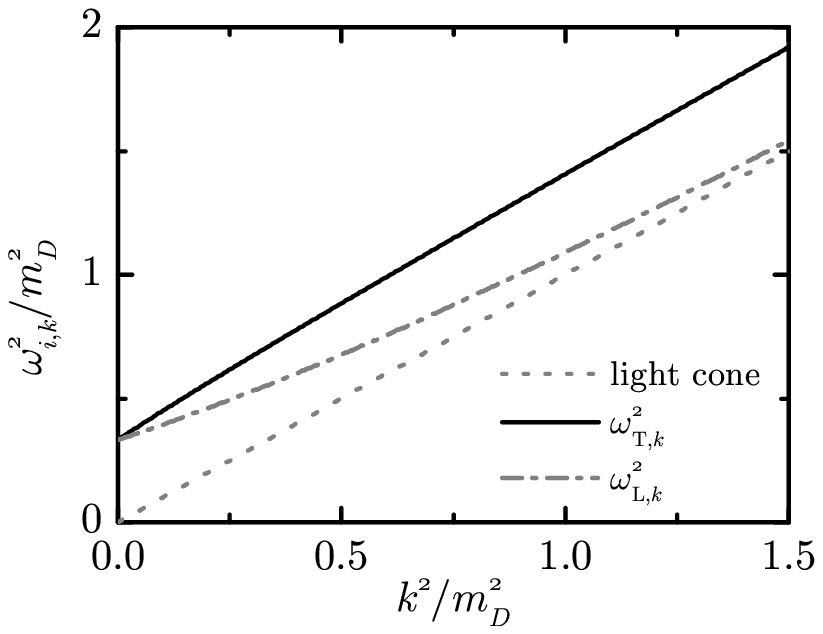}
\par\end{centering}

\caption{The dispersion relations $\omega_{\text{T},k}$ of transverse and
$\omega_{\text{L},k}$ of longitudinal gluon modes scaled by the Debye
screening mass are shown as functions of the momentum $k$ scaled
by the Debye screening mass in linear (left) and quadratic (right)
scales.\label{fig:disp gluons}}

\end{figure}

\begin{figure}[t]
\noindent \begin{centering}
\includegraphics[scale=0.7]{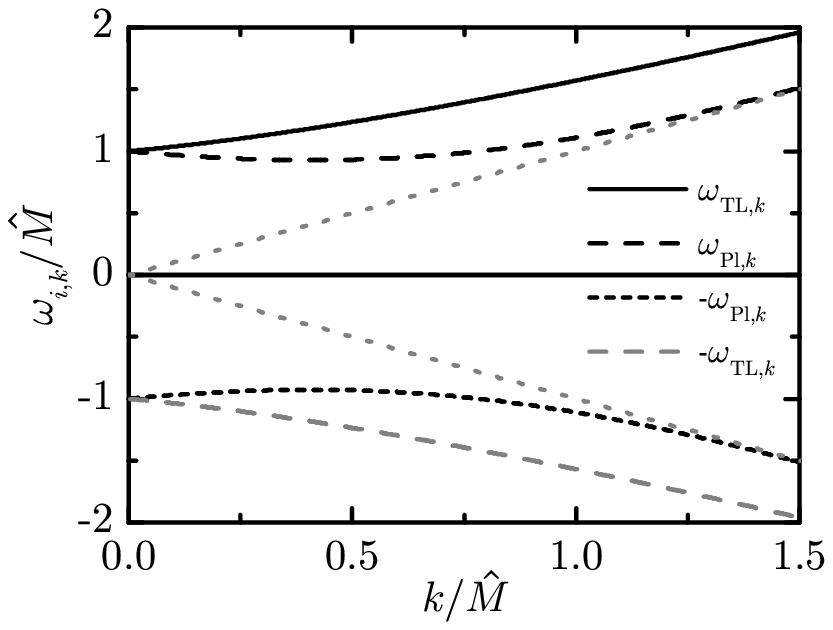}~~~~~~~~\includegraphics[scale=0.7]{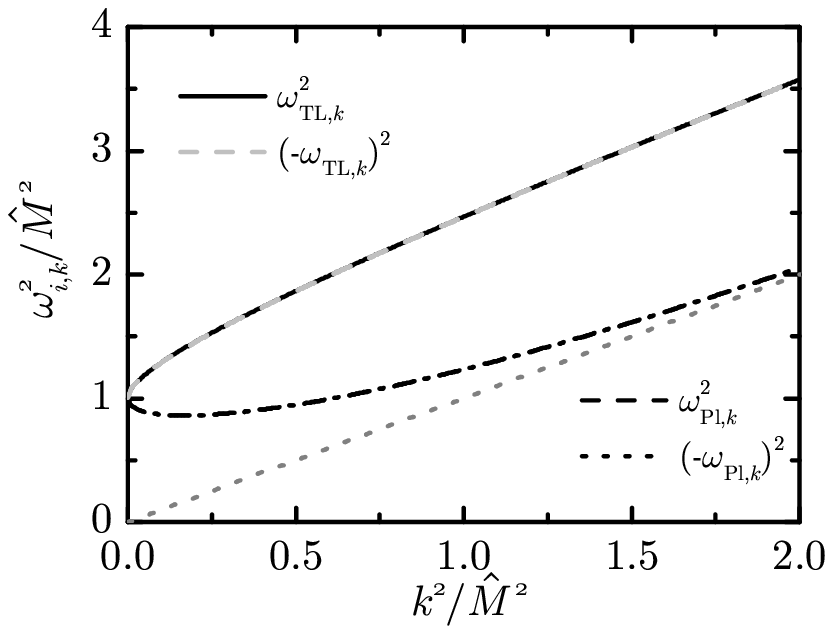}
\par\end{centering}

\caption{The dispersion relations $\omega_{i,k}$ of quarks (solid black),
antiquarks (dashed grey), plasminos (black dashes) and antiplasminos
(black points) scaled by the fermionic mass parameter are shown as
functions of the momentum $k$ scaled by the fermionic mass parameter
in linear (left) and quadratic (right) scales. \label{fig:disp quarks}}

\end{figure}

The evolution of the zeros of the real part of the inverse retarded
propagators as a function of the momentum $k$ gives the dispersion
relations $\omega_{i,k}$. These dispersion relations cannot be expressed
as analytic functions $\omega(k)$ in closed form. $\R D_{i}^{-1}(\omega,\, k,\,\Pi_{i}(\omega,k))=0$
and $\R S_{i}^{-1}(\omega,\, k,\,\Sigma_{i}(\omega,k))=0$ lead to
transcendental equations and have to be solved numerically. The results
are shown in Figures \ref{fig:disp gluons} and \ref{fig:disp quarks}.
Due to the above parity property quarks and antiquarks obey identical
dispersion relations up to the sign, as do plasminos and antiplasminos.%

\subsection{2-loop QCD entropy}

Given the explicit form of the HTL self-energies and the respective
propagators, we evaluate the remaining traces $\text{tr}$ in eq.~(\ref{eq:Omegafinal}).
Assuming equal masses for $u$ and $d$ quarks and zero chemical potential
of strange quarks the isospin chemical potential $\mu_{I}=(\mu_{u}-\mu_{d})/2$
is supposed to vanish at zero net charge. Therefore, there is only
one independent chemical potential $\mu=\mu_{q}=\mu_{B}/3$, where
$\mu_{B}$ is the baryo-chemical potential. As a consequence the flavor
trace gives equal contributions of light quarks to the thermodynamic
potential (i.e. a factor $N_{q}$). The contribution of the strange
quark flavor is here supposed to be equal to the light quark contribution
up to a substitution of $\mu\rightarrow0$ and $N_{q}\rightarrow N_{s}$
in the following formulae.

Taking the trace in Minkowski space, the gluonic part decomposes into
three contributions for one longitudinal and two (equivalent) transverse
polarizations, while the quark contribution becomes the sum of the
normal and the abnormal quark branch (positive and negative chirality
over helicity ratio, respectively) when taking the Dirac trace. The
remaining traces only give overall factors: the color trace $(N_{c}^{2}-1)$
for the gluons and $N_{c}$ for the quarks, and the spin traces for
quarks an additional $2$. Defining the prefactors $d_{g}=N_{c}^{2}-1$,
$d_{q}=2N_{c}N_{q}$ and $d_{s}=2N_{c}N_{s}$ and introducing the
abbreviation $\int_{\dnk}=\int\dnk/(2\pi)^{4}$ the HTL grand canonical
potential then reads\begin{eqnarray}
\frac{\Omega}{V} & = & d_{g}\int_{\dvk}\!\!\nb\,\Big\{2\I\!\left(\ln\DT^{-1}-\DT\PiT\right)+\I\!\left(\ln\left(-\DL^{-1}\right)+\DL\PiL\right)\!\Big\}\\
 &  & +2\sum_{i=q,s}d_{i}\int_{\dvk}\!\!\nf\,\Big\{\I\!\left(\ln S_{+}^{-1}-S_{+}\Sigma_{+}\right)+\I\!\left(\ln\left(-S_{-}^{-1}\right)+S_{-}\Sigma_{-}\right)\!\Big\}-\frac{T}{V}\Gamma_{2}.\nonumber \end{eqnarray}

Differentiating the thermodynamic potential with respect to the temperature
at constant chemical potential gives the entropy. In contrast to the
pressure, which is influenced by vacuum fluctuations, the entropy
is sensitive to thermal excitations and therefore manifestly ultraviolet
(UV) finite. As such, it is ideally suited to investigate the properties
of the QGP \cite{BIR01}.

Due to the stationarity of the thermodynamic potential with respect
to the full propagators, $\delta\Omega/\delta D=0$, only the derivatives
of the statistical distribution functions contribute. Using $\I(\DT\PiT)=\R\DT\I\PiT+\I\DT\R\PiT$,
the entropy density can be written as $s:=-V^{-1}\left.\partial\Omega/\partial T\right|_{\mu}=s_{g,\text{T}}+s_{g,\text{L}}+\sum_{q,s}(s_{i,+}+s_{i,-})$
with\begin{eqnarray}
s_{g,\text{T}} & = & -2d_{g}\int_{\dvk}\frac{\partial\nb(\omega)}{\partial T}\Big\{\mbox{Im}\ln\left(+\DT^{-1}\right)-\mbox{Re}\DT\mbox{Im}\PiT\Big\},\label{eq:sgT ImLn}\\
s_{g,\text{L}} & = & -\,\,\, d_{g}\int_{\dvk}\frac{\partial\nb(\omega)}{\partial T}\Big\{\mbox{Im}\ln\left(-\DL^{-1}\right)+\mbox{Re}\DL\mbox{Im}\PiL\Big\},\label{eq:sgL ImLn}\\
s_{q/s,\pm} & = & -2d_{q/s}\!\int_{\dvk}\!\!\frac{\partial\nf(\omega)}{\partial T}\Big\{\mbox{Im}\ln\left(\pm S_{\pm}^{-1}\right)\mp\mbox{Re}S_{\pm}\mbox{Im}\Sigma_{\pm}\Big\},\label{eq:sq ImLn}\end{eqnarray}
each describing the entropy density of one quasiparticle species in
the absence of the others. An interaction (correlation) entropy density
contribution would contain terms of the form $\I\DT\R\PiT$ and the
derivative of $\Gamma_{2}T$ with respect to the temperature. However,
at 2-loop order these terms exactly cancel each other \cite{BIR01}.
In fact, this seems to be a generic, topological feature \cite{CP75}
which has explicitly been proven for QED \cite{VB98} and $\Phi^{4}$
theory \cite{Pes01} too.

We now focus on the terms $\I\ln(\pm D_{\text{T,L}}^{-1})$ and $\I\ln(\pm S_{\pm}^{-1})$,
which can be written as\begin{eqnarray}
\I\ln\DT^{-1} & = & \arctan\left(\frac{\I\DT^{-1}}{\R\DT^{-1}}\right)+\pi\varepsilon(\I\DT^{-1})\Theta\!\left(-\R\DT^{-1}\right),\label{eq:ImLn+DTm1}\\
\I\ln\left(-\DL^{-1}\right) & = & \arctan\left(\frac{\I\DL^{-1}}{\R\DL^{-1}}\right)-\pi\varepsilon(\I\DL^{-1})\Theta\!\left(+\R\DL^{-1}\right).\label{eq:ImLn-DLm1}\end{eqnarray}
Similar expressions apply for the two quark propagators: one has to
substitute $S_{+}^{-1}$ for $\DT^{-1}$ in (\ref{eq:ImLn+DTm1})
and $S_{-}^{-1}$ for $\DL^{-1}$ in (\ref{eq:ImLn-DLm1}).

From the properties of the imaginary parts of the self-energies discussed
above, we find $\e(\I D_{i}^{-1}(\omega))=-\e(\omega)$ for the gluons
and $\e(\I S_{\pm}(\omega))\equiv-1$ for the normal and abnormal
quark branches. We end up with\begin{eqnarray}
s_{g,\text{T}} & = & +2d_{g}\int_{\dvk}\frac{\partial\nb}{\partial T}\Big\{\pi\varepsilon(\omega)\Theta\!\left(-\R\DT^{-1}\right)-\arctan\frac{\I\PiT}{\R\DT^{\text{-}1}}+\mbox{Re}\DT\mbox{Im}\PiT\Big\},\label{eq:si eq theta arctan reim gT}\\
s_{g,\text{L}} & = & -\,\,\, d_{g}\int_{\dvk}\frac{\partial\nb}{\partial T}\Big\{\pi\varepsilon(\omega)\Theta\!\left(+\R\DL^{-1}\right)-\arctan\frac{\I\PiL}{\R\DL^{\text{-}1}}+\mbox{Re}\DL\mbox{Im}\PiL\Big\},\label{eq:si eq theta arctan reim gL}\\
s_{q/s,\pm} & = & \pm2d_{q/s}\!\int_{\dvk}\frac{\partial\nf}{\partial T}\Big\{\quad\,\,\,\pi\Theta\!\left(\mp\R S_{\pm}^{-1}\right)-\arctan\frac{\I\Sigma_{\pm}}{\R S_{\pm}^{\text{-}1}}\,+\mbox{Re}S_{\pm}\mbox{Im}\Sigma_{\pm}\Big\}.\label{eq:si eq theta arctan reim q}\end{eqnarray}
The partial entropy densities (\ref{eq:si eq theta arctan reim gT})-(\ref{eq:si eq theta arctan reim q})
and, therefore the whole entropy density expression, are independent
of possible renormalization factors. As required, the expression is
also explicitly UV finite, as the derivatives of the distribution
functions soften the UV behavior. The terms $\pi\Theta(\ldots)$ represent
the quasiparticle contributions to the entropy, while the terms containing
the imaginary parts of the self-energies are related to damping effects
and quasiparticle widths. In the case of HTL self-energies, Landau
damping is contained within the latter terms.

The quark entropy density $s_{q/s}=s_{q/s,+}+s_{q/s,-}$ can be simplified
by utilizing the parity properties for quark propagators and self-energies.
Introducing the distribution function of antiparticles $\nf^{A}=(e^{\beta(\omega+\mu)}+1)^{-1}$
with $\partial\nf(-\omega)/\partial\omega=-\partial\nf^{A}(\omega)/\partial\omega$
and substituting $\omega\rightarrow-\omega$ within $s_{q,-}$, we
find\begin{equation}
s_{q/s}=2d_{q/s}\int_{\dvk}\left(\frac{\partial\nf}{\partial T}\!+\!\frac{\partial\nf^{A}}{\partial T}\right)\left\{ \pi\Theta\!\left(\text{-}\R S_{+}^{-1}\right)-\arctan\!\left(\frac{\I\Sigma_{+}}{\R S_{+}^{\text{-}1}}\!\right)+\mbox{Re}S_{+}\mbox{Im}\Sigma_{+}\right\} .\label{eq:sq combined}\end{equation}
Regarding the quasiparticle pole term $\pi\Theta(\text{-}\R S_{+}^{-1})$,
the energy integration from $-\infty$ to $0$ gives the (anti)plasmino
contribution, while the integration from $0$ to $+\infty$ delivers
the contributions of the (anti)particles to the entropy density. Isolating
both parts of the spectrum by applying the parity properties once
more gives the explicit expressions\begin{eqnarray}
s_{q/s,\text{TL}} & \!\!=\!\! & \,\,\,\,\,2d_{q/s}\int_{\ddk}\int\limits _{0}^{\infty}\!\frac{\dw}{2\pi}\,(.)\left\{ \pi\Theta\!\left(\text{-}\R S_{+}^{-1}\right)-\arctan\!\left(\frac{\I\Sigma_{+}}{\R S_{+}^{\text{-}1}}\!\right)+\mbox{Re}S_{+}\mbox{Im}\Sigma_{+}\!\right\} ,\label{eq:sq TL-Pl split}\\
s_{q/s,\text{Pl}} & \!\!=\!\! & -2d_{q/s}\int_{\ddk}\int\limits _{0}^{\infty}\!\frac{\dw}{2\pi}\,(.)\left\{ \pi\Theta\!\left(\,\,\R S_{-}^{-1}\right)-\arctan\!\left(\frac{\I\Sigma_{-}}{\R S_{-}^{\text{-}1}}\!\right)+\mbox{Re}S_{-}\mbox{Im}\Sigma_{-}\!\right\} ,\label{eq:sq TL-Pl split Pl}\end{eqnarray}
where the sum of the derivatives of the distribution functions is
abbreviated by the parentheses $(.)$. While this separation seems
straightforward, it has to be handled with care as the Landau damping
term within the quark self-energies $\Sigma_{\pm}$ (see the imaginary
parts in Figure \ref{fig:quark se}) can, in general, not be separated
into quark and plasmino contributions in this simple way.

\subsection{The full HTL QPM}

Since the entropy density of the quark-gluon plasma for 2-loop QCD
is the sum of the single quasiparticle entropy density contributions,
it can be considered as mixture of non-interacting ideal quasiparticle
gases. It is natural to assume that the pressure, which follows from
the entropy density by integration, consists of single partial pressures,
too. Therefore, we use the ansatz $p=p_{g,\text{T}}+p_{g,\text{L}}+\sum_{i=q,s}p_{i}-B(\PiT,\PiL,\Sigma_{\pm})$
for the pressure, where $B$ is chosen appropriately to ensure thermodynamic
consistency. The ansatz has to satisfy $s_{i}=\partial p_{i}/\partial T|_{\mu}$
which leads to\begin{eqnarray}
p_{g,\text{T}} & = & +2d_{g}\int_{\dvk}\nb\Big\{\pi\varepsilon(\omega)\Theta\!\left(-\R\DT^{-1}\right)-\arctan\frac{\I\PiT}{\R\DT^{\text{-}1}}+\mbox{Re}\DT\mbox{Im}\PiT\Big\},\\
p_{g,\text{L}} & = & -\,\,\, d_{g}\int_{\dvk}\nb\Big\{\pi\varepsilon(\omega)\Theta\!\left(+\R\DL^{-1}\right)-\arctan\frac{\I\PiL}{\R\DL^{\text{-}1}}+\mbox{Re}\DL\mbox{Im}\PiL\,\Big\},\\
p_{q/\! s} & = & 2d_{q/\! s}\!\int_{\dvk}\!\!\left(\nf\!+\!\nf^{A}\right)\!\Big\{\pi\Theta\!\left(-\R S_{+}^{-1}\right)-\arctan\frac{\I\Sigma_{+}}{\R S_{+}^{\text{-}1}}\,+\mbox{Re}S_{+}\mbox{Im}\Sigma_{+}\Big\},\end{eqnarray}
where the integrability condition $\partial B/\partial\Pi_{i}=\partial p/\partial\Pi_{i}$
has to be fulfilled for every quasiparticle species $i$. Thus $B$
ensures the stationarity of the thermodynamic potential under functional
variation with respect to the self-energies \cite{GY95}. Note that
the plasma frequency within the $s$-quark pressure differs from the
plasma frequency within $p_{q}$ as $\mu_{s}=0$.

The pressure fully defines the model. The particle density follows
by differentiation of the pressure with respect to the chemical potential
at constant temperature. The Bose-Einstein distribution function $\nb$
does not depend on $\mu_{g}$ and strange quarks are included into
the model with manifest zero net particle density, therefore $n_{g,\text{T}}=n_{g,\text{L}}=n_{s}=0$
. Due to the integrability condition, the terms containing the derivatives
of the self-energies with respect to $\mu$ vanish, so that\begin{equation}
n\!=\! n_{q}\!\!=\!2d_{q}\!\!\int_{\dvk}\!\!\left(\frac{\partial\nf}{\partial\mu}\!+\!\frac{\partial\nf^{A}}{\partial\mu}\right)\!\!\left\{ \pi\Theta\!\left(\text{-}\R S_{+}^{-1}\right)\!-\!\arctan\frac{\I\Sigma_{+}}{\R S_{+}^{\text{-}1}}\!+\!\mbox{Re}S_{+}\mbox{Im}\Sigma_{+}\right\} ,\label{eq:fullHTL nq}\end{equation}
thus $n_{q}(\mu\rightarrow0)\rightarrow0$.

\subsection{Effective coupling}

Obviously, 2-loop QCD is only a crude approximation of the full theory.
In order to accommodate non-perturbative effects in the quasiparticle
model, we introduce some flexibility by parameterizing the QCD coupling
constant $g^{2}$ in a phenomenologically motivated way. The truncated
2-loop running QCD coupling $g^{2}$ is given by $g^{2}(x)=16\pi^{2}(\beta_{0}\ln(x))^{-1}\,(1-2\beta_{1}\ln[\ln(x)]\,(\beta_{0}^{2}\ln(x))^{-1})$
\cite{PDG06}, where $\beta_{0}=11N_{c}/3-2N_{f}/3$, $\beta_{1}=51-19N_{f}/3$
and $N_{f}=N_{q}+N_{s}$. It depends on the ratio $x=\bar{\mu}^{2}/\Lambda^{2}$
of the renormalization scale $\bar{\mu}$ and the QCD scale parameter
$\Lambda$. A term involving $\ln^{-2}(\bar{\mu}^{2}/\Lambda^{2})$
which is only a small correction for $\bar{\mu}^{2}\approx\Lambda^{2}$
was neglected.

The renormalization scale is usually taken to be the first Matsubara
frequency $2\pi T$, while the latter one is just a parameter to be
adjusted using experimental data. Introducing the pseudocritical temperature
of QCD matter at vanishing net baryon density $T_{c}$ and substituting
$\Lambda\rightarrow2\pi T_{c}/\lambda$, the ratio $\bar{\mu}/\Lambda$
becomes $\lambda T/T_{c}$. In order to avoid the Landau pole of $g^{2}(T/T_{c})$
at $T_{c}$ a temperature shift with parameter $T_{s}$ is introduced,
replacing $x$ by $\xi=\lambda(T-T_{s})/T_{c}$. The result \begin{equation}
G^{2}(T\geq T_{c},\mu=0)=\frac{16\pi^{2}}{\beta_{0}\ln\xi^{2}}\left(1-\frac{2\beta_{1}}{\beta_{0}^{2}}\frac{\ln\left[\ln\xi^{2}\right]}{\ln\xi^{2}}\right)\label{eq:eff coupling}\end{equation}
is our \emph{effective coupling}. Within the plasma phase for temperatures
$T>T_{c}/\lambda+T_{s}$ it is well-behaved; however, at some point
within the hadronic phase, i.e.~below $T_{c}$, an infrared (IR)
divergence does occur. In order to prevent this divergence a phenomenological
infrared cutoff for $G^{2}$ can be applied. See \cite{Blu04a} for
details.

\subsection{Adjustment to lattice calculations}

The two QPM parameters $\lambda$ and $T_{s}$ have to be adjusted
to results of numerical first-principle QCD calculations dubbed lattice
data. Most of the past work on the QPM has been tested against lattice
data from \cite{KLP00} for rather large and temperature dependent
lattice restmasses of $m_{q}=0.4T$ and $m_{s}=1.0T$ compared to
the physical quark masses $m_{u,d}\sim10\,\MeV$ and $m_{s}\sim90-150\,\MeV$
\cite{PDG06}. Recently, new lattice data has become available \cite{Kar07},
which relies on lattice restmasses much closer to the physical quark
masses and which is used in this work.

Also, lattice calculations are performed on a finite lattice, while
our quasiparticle model is formulated in the thermodynamic limit,
i.e.~aimed at describing a spatially infinite plasma. In order to
compare our model with lattice data, the proper continuum extrapolation
of the latter one is required. A safe continuum extrapolation on the
lattice is a fairly demanding work. Therefore, various estimates have
been applied, e.g.~simply scaling the lattice results by a factor
being strictly valid only for asymptotically high temperatures or
for the non-interacting limit. To account for a possible deficit of
such rough continuum estimates of the lattice data we introduce an
ad hoc scaling factor $d_{\text{lat}}$ which turns out to be nearly
unity.

\subsection{Nonzero chemical potential}

The parametrization of $G^{2}$ in eq.~(\ref{eq:eff coupling}) is
valid for $\mu=0$ only. However, it is possible to use the thermodynamic
consistency of the QPM to map the results at zero chemical potential
into the $T$-$\mu$ plane \cite{Pes00}. Specifically, this means
to impose the Maxwell relation $\partial s/\partial\mu|_{T}=\partial n/\partial T|_{\mu}$
on the thermodynamic quantities. Ordering the terms with respect to
the partial derivatives of the effective coupling gives an elliptic
quasilinear partial differential equation\begin{equation}
a_{T}\frac{\partial G^{2}}{\partial T}+a_{\mu}\frac{\partial G^{2}}{\partial\mu}=b,\label{eq:floweq}\end{equation}
named hereafter flow equation, with the coefficients $a_{T}$, $a_{\mu}$
and $b$ depending on $T$ and $\mu$ explicitly and, via the self-energies,
implicitly. It is solved by the method of characteristics by introducing
a curve parameter $x$, assuming that $T=T(x)$, $\mu=\mu(x)$ and
$G^{2}=G^{2}(x)$. Subsequently, the comparison of $G_{,x}^{2}=G_{,T}^{2}T_{,x}+G_{,\mu}^{2}\mu_{,x}$
with the flow equation gives a system of three linear, coupled ordinary
differential equations: $G_{,x}^{2}=-b$, $T_{,x}=-a_{T}$ and $\mu_{,x}=-a_{\mu}$
which can be solved using standard numerical methods. The initial
condition for the flow equation is the effective coupling at $\mu=0$,
with model parameters fixed by comparison of the entropy density with
lattice results.

\section{The effective QPM}

\label{sec:eQP}

Assuming that transversal gluons and quark particle excitations propagate
predominantly on mass shells the full HTL QPM can be significantly
simplified as it implies explicit (asymptotic) dispersion relations
$\omega_{i}(k)$ of the form $\omega_{i}^{2}(k)=k^{2}+m_{i,\infty}^{2}$
as approximations to the full, implicit ones \cite{Pes02}. The $m_{i,\infty}$
terms depend neither on energy nor momentum and are therefore called
asymptotic (thermal) masses. In order to adjust the eQP to lattice
data, they can be modified to accomodate lattice restmasses $m_{i}$
using a prescription from \cite{Pis93}: $m_{i,\infty}^{2}\rightarrow m_{i}^{2}+2m_{i}m_{i,\infty}+2m_{i,\infty}^{2}$.
Additionally neglecting Landau damping, i.e.~assuming vanishing imaginary
parts of the self-energies, and contributions from collective excitations,
i.e.~plasmons and (anti)plasminos, leads to the eQP. As collective
excitations are exponentially suppressed%
\footnote{That is after calculating the propagators using Dyson's relation from
the HTL self-energies, the residues of the poles in the spectral density
of both plasmon and (anti)plasmino propagators vanish exponentially
for momenta $k\sim T,\mu$, which give the dominant main contributions
to thermodynamic integrals.%
} and the effect of Landau damping is small at vanishing chemical potential
the eQP seemed to be sufficient.

However, as previous studies of the flow equation \cite{Pes02,Blu04a}
have shown the characteristic curves emerging at $T\approx T_{c}$
cross each other in some region of finite values of $\mu$ for parameters
adjusted to lattice QCD results (cf.~dashed lines in Figure \ref{fig:HTL characteristics}
below). This unfortunate feature prevents an unambiguous extrapolation
of thermodynamic quantities into the full $T$-$\mu$-plane. Romatschke
\cite{RR03} has shown for the $2$ flavor case that these crossings
can be avoided by using the full HTL model. We are going to extend
his line of arguing to the physically interesting case of $2+1$ flavors.

\section{Investigation of the full HTL QPM\label{sec:fullHTL}}

\subsection{Adjustment to lattice data at $\mu=0$}

While the eQP is able to accommodate arbitrary lattice restmasses
by means of a modified asymptotic dispersion relation, the full HTL
model relies on the HTL dispersion relations and thus massless particles.
We assume here that the employed lattice restmasses in \cite{Kar07}
are sufficiently small to be absorbed in suitably adjusted parameters.

\begin{figure}
\begin{centering}
\includegraphics[scale=0.7]{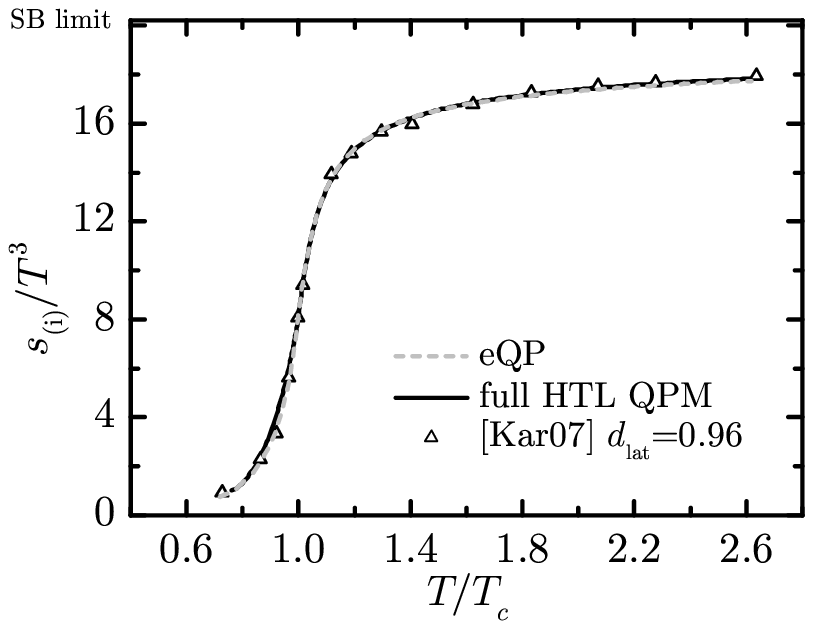}~~~~~~~~\includegraphics[scale=0.7]{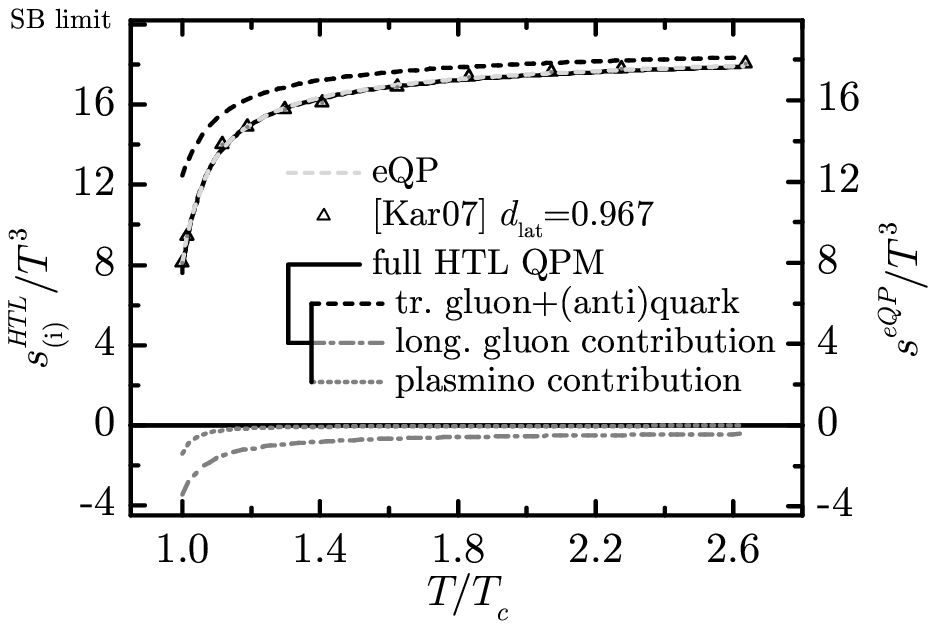}
\par\end{centering}

\caption{The scaled entropy densities $s/T^{3}$ of the full HTL QPM with quadratic
IR regulator (solid black lines; $T_{s}=0.728T_{c}$ and $\lambda=6.10$)
and the eQP (grey dashed lines; $T_{s}=0.752T_{c}$ and $\lambda=6.26$)
adjusted to lattice data for $N_{f}=2+1$ from \cite{Kar07} with
$d_{\text{lat}}=0.96$ are shown as functions of the scaled temperature
$T/T_{c}$. The adjustment quality of the full HTL QPM to lattice
data is indistinguishable from the eQP. The single contributions to
$s^{HTL}$, including their respective LD contributions, are given
in the right figure (dashed black: transversal gluons+(anti)quarks,
dash-dotted: longitudinal gluons, dotted grey: (anti)plasminos). \label{fig:fullHTL fit quad}\label{fig:sHTL contributions}}

\end{figure}

For $T>T_{c}$ and $N_{f}=2$ the full HTL model has been shown to
give a description of lattice data being equally well as the eQP \cite{Rom04}.
For $N_{f}=2+1$ flavors we meet a similar situation: The full HTL
QPM describes the lattice QCD data \cite{Kar07} as good as the eQP
model (see Figure \ref{fig:fullHTL fit quad}, left). The extension
to $T<T_{c}$, on the other hand, is not straightforward. Instead
of a linear IR regulator \cite{Blu04b}, it is necessary to use a
quadratic parametrization of the effective coupling in order to achieve
agreement with lattice data also below the pseudocritical temperature,
which we consider here as mere parametrisation of the lattice data.%

When evaluating the individual contributions to the entropy density
of the full HTL QPM we find the entropy density contributions of longitudinal
gluon $s_{g,\text{L}}$ (eq.~(\ref{eq:si eq theta arctan reim gL}))
and (anti)plasminos $s_{q,\text{Pl}}$ (eq.~(\ref{eq:sq TL-Pl split Pl}))
to be negative. This is due to the fact that both represent collective
phenomena of the QGP resulting in correlations not present in a noninteracting
medium. As a consequence, the transverse gluon and (anti)quark entropy
density contributions have to increase in comparison to the eQP in
order for the sum of the partial entropy densities to describe the
same lattice data as the eQP (see Figure \ref{fig:sHTL contributions},
right). This not only allows for a pure quasiparticle entropy contribution
much closer to the Stefan-Boltzmann limit than in the eQP but also
proves to have a positive impact on the extension to nonzero chemical
potential.

\subsection{Solution of the flow equation}

\begin{figure}
\begin{centering}
\includegraphics[bb=0bp 0bp 253bp 185bp,clip,scale=0.7]{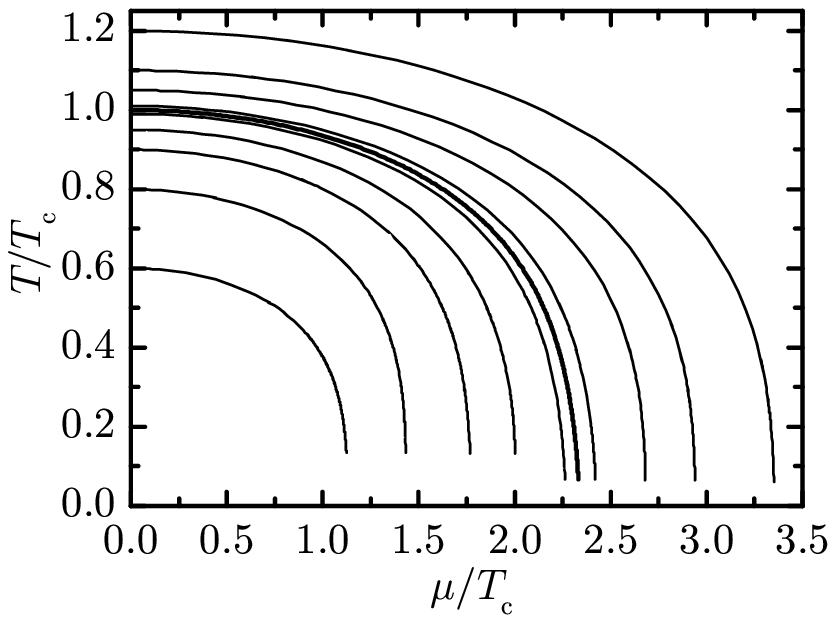}~~~~~~~~\includegraphics[bb=0bp 0bp 253bp 185bp,clip,scale=0.7]{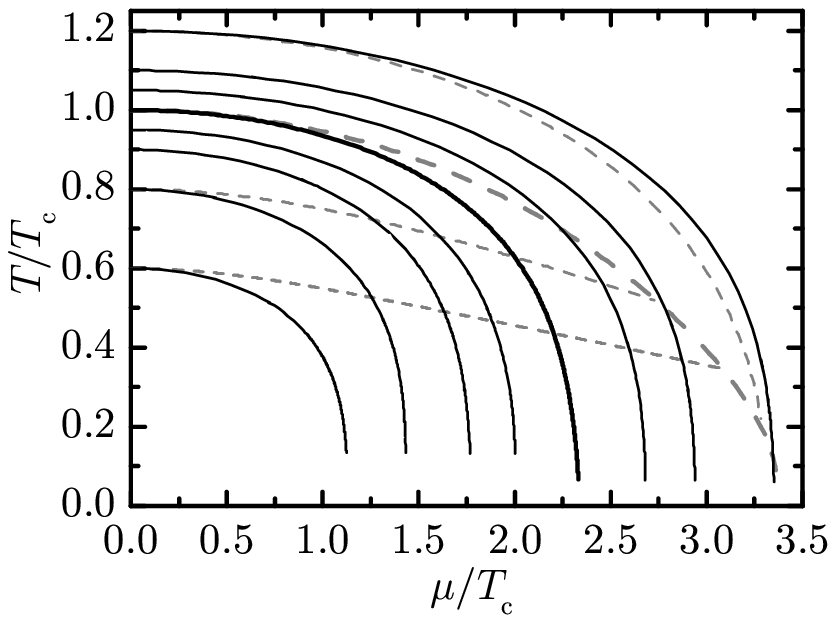}
\par\end{centering}

\caption{The solid curves in both graphs are several characteristics of the
full HTL flow equation for $2+1$ quark flavors using parameters from
the adjustment of the full HTL QPM to lattice data from \cite{Kar07}
shown in Figure \ref{fig:fullHTL fit quad}. The characteristic curve
emerging from $T_{c}$ is depicted as bold solid line. All crossings
have disappeared (left panel). For a comparison, the right panel shows
the characteristics of the eQP flow equation using the parameters
of the eQP adjustment to the same lattice QCD data (dashed curves).}
\label{fig:HTL characteristics}
\end{figure}

Solving the flow equation (\ref{eq:floweq}) with coefficients listed
in Appendix \ref{app:flow eq coeff} the characteristics are found
to be well-behaving, as can be seen in Figure \ref{fig:HTL characteristics}.
Also a stronger curvature of the full HTL characteristics compared
to the eQP characteristics (shown as dashed lines in the right panel)
is observable.

To explain the disappearance of the ambiguities caused by crossing
characteristics in the eQP we mention that the crossings appear due
to the effective coupling $G^{2}$ being too large near the pseudocritical
temperature \cite{Blu04}. Since the entropy density increases with
decreasing mass parameters $m_{D}^{2}$ and $\hat{M}^{2}$ (which
are proportional to $G^{2}T^{2}$ at $\mu=0$) the crossings would
therefore disappear for a larger eQP entropy density. One way to allow
for a larger eQP entropy density is to take into account collective
modes. As medium effects indicate correlations between the constituents
of the eQP plasma, including them causes a decrease of overall entropy
density. Consequently, the eQP parameters have to change in order
to still describe the same lattice data, causing the entropy density
to increase. With the resulting decrease of the effective coupling
$G^{2}$ the crossings partially disappear.

However, the different parametrization at $\mu=0$ alone cannot account
for the complete absence of ambiguities for the full HTL model. Instead,
the influence of collective modes and Landau damping on the flow equation
has to be examined. We therefore calculate the characteristics of
the full HTL model respectively disregarding terms stemming from these
contributions. While neglecting plasmon/(anti)plasminos terms from
the coefficients $a_{T}$, $a_{\mu}$ and $b$ (see Appendix \ref{app:flow eq coeff})
but keeping the Landau damping contributions leads to deformed characteristics
meeting $T=0$ at smaller $\mu$ and no crossings appear. Hence, it
is neither the plasmon nor the plasmino term which accounts for the
vanishing crossings. However, neglecting the Landau damping terms
immediately leads to crossing characteristics. Therefore, both collective
excitations (in order to obtain a reasonably small coupling $G^{2}$)
and Landau damping (in order to ultimately remove the crossings) are
necessary to obtain a flow equation with unique solutions. Using this
flow equation, it is possible to extrapolate the equation of state
from lattice QCD at $\mu=0$ towards $T=0$.%

\section{Conclusion}

\label{sec:conclusion}

The mapping of a previous quasiparticle model (eQP) into the $T$-$\mu$
plane was plagued by crossing characteristics. It is shown here for
the 2+1 flavor case that, if using the full HTL model, these crossings
disappear. Collective modes (longitudinal gluon and (anti)quark hole
excitations, i.e.~plasmons and (anti)plasminos respectively) as well
as Landau damping of the collisionless quasiparticle plasma, both
neglected hitherto in the eQP, need to be taken into account to avoid
the ambiguities.

With the problem of crossing characteristics solved, one can proceed
to derive an equation of state, following from the full HTL QPM, especially
for the cold and dense quark-gluon plasma of interest in future heavy
ion collision experiments or for the simulation of possible quark
stars.

\begin{acknowledgement}
\emph{Acknowledgment:} R.S. would like to thank the organizers of
the \emph{Zim\'anyi 75 Memorial Workshop} for the invitation to present
his results at this very inspiring workshop.
\end{acknowledgement}
\appendix

\section{Coefficients of the flow equation\label{app:flow eq coeff}}

For the reader's convenience and to extend the results in \cite{Rom04}
to $N_{f}=2+1$ flavors the full HTL flow equation is presented. To
calculate the Maxwell relation, the derivatives $\partial s_{g}/\partial\mu=A_{g}\partial m_{D}^{2}/\partial\mu$,
$(\partial s_{q}/\partial\mu)_{impl.}=A_{q}\partial\hat{M}^{2}/\partial\mu$,
$(\partial s_{s}/\partial\mu)_{impl.}=A_{s}\partial\hat{M}^{2}/\partial\mu$
and $(\partial n_{q}/\partial T)_{impl.}=A_{n}\partial\hat{M}^{2}/\partial T$
are necessary. The explicit derivatives cancel within the Maxwell
relation due to Schwarz's Theorem. We find\begin{eqnarray}
A_{g} & = & \frac{d_{g}}{2\pi^{3}m_{D}^{2}}\int\limits _{0}^{\infty}\dk\, k^{2}\Bigg(\int\limits _{0}^{k}\dw\left[\frac{\partial\nb}{\partial T}\frac{4(\omega^{2}-k^{2})\I^{3}\PiT}{(\R^{2}\DT^{-1}\!+\!\I^{2}\PiT)^{2}}\!-\!\frac{\partial\nb}{\partial T}\frac{2k^{2}\I^{3}\PiL}{(\R^{2}\DL^{-1}\!+\!\I^{2}\PiL)^{2}}\right]\label{eq:dsgdmu HTL}\\
 &  & \quad\quad\quad\quad\quad\quad\quad\quad-\,\,\pi\,\frac{\omega_{\text{T},k}(\omega_{\text{T},k}^{2}-k^{2})^{2}}{|(\omega_{\text{T},k}^{2}\!-\! k^{2})^{2}\!-\! m_{D}^{2}\omega_{\text{T},k}^{2}|}\left.\frac{\partial\nb}{\partial T}\right|_{\omega_{\text{T},k}}\!\!\!\!\!\!\!\!-\,\pi\,\frac{\omega_{\text{L},k}(\omega_{\text{L},k}^{2}-k^{2})}{|\omega_{\text{L},k}^{2}\!-\! k^{2}\!-\! m_{D}^{2}|}\left.\frac{\partial\nb}{\partial T}\right|_{\omega_{\text{L},k}}\Bigg)\nonumber \end{eqnarray}
\begin{eqnarray}
A_{q} & = & \frac{d_{q}}{2\pi^{3}\hat{M}^{2}}\int\limits _{0}^{\infty}\dk\, k^{2}\Bigg(\int\limits _{-k}^{k}\dw\left[{\cal N}_{T}\frac{2(\omega-k)\I^{3}\Sigma_{+}}{(\R^{2}S_{+}^{-1}\!+\!\I^{2}\Sigma_{+})^{2}}\right]\label{eq:dsqdmu HTL}\\
 &  & \quad\quad\quad\quad\quad\quad\quad\,\,\,\,-\pi\frac{\omega_{\text{TL},k}^{2}\!\!-k^{2}}{2\hat{M}^{2}}(\omega_{\text{TL},k}-k)\left.{\cal N}_{T}\right|_{\omega_{\text{TL},k}}-\pi\frac{\omega_{\text{Pl},k}^{2}\!\!-k^{2}}{2\hat{M}^{2}}(\omega_{\text{Pl},k}+k)\left.{\cal N}_{T}\right|_{\omega_{\text{Pl},k}}\Bigg)\nonumber \end{eqnarray}
\begin{eqnarray}
A_{n} & = & \frac{d_{q}}{2\pi^{3}\hat{M}^{2}}\int\limits _{0}^{\infty}\dk\, k^{2}\Bigg(\int\limits _{-k}^{k}\dw\left[{\cal N}_{\mu}\frac{2(\omega-k)\I^{3}\Sigma_{+}}{(\R^{2}S_{+}^{-1}\!+\!\I^{2}\Sigma_{+})^{2}}\right]\label{eq:dnqdT HTL}\\
\, & \, & \quad\quad\quad\quad\quad\quad\quad\,\,\,\,-\pi\frac{\omega_{\text{TL},k}^{2}\!\!-k^{2}}{2\hat{M}^{2}}(\omega_{\text{TL},k}-k)\left.{\cal N}_{\mu}\right|_{\omega_{\text{TL},k}}-\pi\frac{\omega_{\text{Pl},k}^{2}\!\!-k^{2}}{2\hat{M}^{2}}(\omega_{\text{Pl},k}+k)\left.{\cal N}_{\mu}\right|_{\omega_{\text{Pl},k}}\Bigg)\nonumber \end{eqnarray}
with abbreviations ${\cal N}_{T}:=\partial/\partial T(\nf+\nf^{A})$
and ${\cal N}_{\mu}:=\partial/\partial\mu(\nf+\nf^{A})$. The derivative
of the strange quark entropy density with respect to the temperature
equals the light quark expression at vanishing chemical potential
with $A_{s}=A_{q}(\mu=0)$ and $\partial\hat{M}_{s}^{2}/\partial\mu=(\partial\hat{M}^{2}/\partial\mu)|_{\mu=0}$.
Imposing the Maxwell relation $\partial s/\partial\mu|_{T}=\partial n/\partial T|_{\mu}$
and employing the prefactors in eq.~(\ref{eq:mDebye and plasma freq})
the coefficients of the flow equation (\ref{eq:floweq}) are given
by\begin{eqnarray}
a_{T} & = & -\frac{N_{c}^{2}-1}{16N_{c}}\left(T^{2}+\frac{\mu^{2}}{\pi^{2}}\right)A_{n},\label{eq:full HTL floweq aT}\\
a_{\mu} & = & \frac{1}{6}\left(\left[2N_{c}+N_{q}+N_{s}\right]T^{2}+\frac{N_{c}N_{q}}{\pi^{2}}\mu^{2}\right)A_{g}\label{eq:full HTL floweq amu}\\
 &  & \quad\quad\quad\quad\quad\quad\,\,\,+\frac{N_{c}^{2}-1}{16N_{c}}\left(T^{2}+\frac{\mu^{2}}{\pi^{2}}\right)A_{q}+\frac{N_{c}^{2}-1}{16N_{c}}T^{2}A_{s},\nonumber \\
b & = & \frac{N_{c}^{2}-1}{8N_{c}}TG^{2}A_{n}-\frac{N_{c}N_{q}}{3\pi^{2}}\mu\, G^{2}A_{g}-\frac{N_{c}^{2}-1}{8N_{c}\pi^{2}}\mu\, G^{2}A_{q}.\label{eq:full HTL floweq b}\end{eqnarray}
These expressions correct a few typos in \cite{Rom04} (see eqs.~(B.1)-(B.5)
therein).

\bibliographystyle{epj}

\end{document}